\documentclass[12pt,a4paper]{article}
\usepackage{graphicx}% Include figure files

\begin{document}
\textwidth=135mm
 \textheight=200mm

\begin{center}{
{\bfseries {Sources of the systematic errors in  measurements  of
$^{214}$Po decay half-life time variations at the Baksan deep
underground  experiments}} \footnote{{\small Talk at The
International Workshop on Prospects of Particle Physics: "Neutrino
Physics and Astrophysics" January 26 - Ferbuary 2, 2014, Valday,
Russia}}} \vskip 5mm E.N.~Alexeyev$^{\dag}$, Yu.M.
Gavrilyuk$^\dag$, A.M.~Gangapshev$^\dag$, V.V.~Kazalov$^\dag$,
V.V.~Kuzminov$^\dag$, S.I.~Panasenko$^{\dag,\ddag}$,
S.S.~Ratkevich$^{\dag,\ddag}$ \vskip 5mm {\small
{\it $^\dag$Baksan Neutrino Observatory INR RAS, Russiaa}} \\
{\small {\it $^\ddag$ V.N.Karazin Kharkiv National University,
Ukraine}}
\\
\end{center}

\vskip 5mm \centerline{\bf Abstract} The design changes of the
Baksan  low-background TAU-1 and TAU-2 set-ups allowed to improve
a sensitivity of $^{214}$Po half-life ($\tau$) measurements up to
the $2.5\cdot10^{-4}$ are described. Different possible sources of
systematic errors influencing on the $\tau$-value are studied. An
annual variation of $^{214}$Po half-life time  measurements with
an amplitude of $A=(6.9 \pm 3) \times 10^{-4}$ and a phase of
$\varphi=93 \pm 10$ days was found in a sequence of the
week-collected $\tau$-values obtained from the TAU-2 data sample
with total duration of 480 days. 24 hours' variation of the
$\tau$-value measurements with an amplitude of
$A=(10.0\pm2.6)\times 10^{-4}$ and phase of $\varphi=1\pm0.5$ hours
was found in a solar day 1 hour step $\tau$-value sequence formed
from the same data sample. It was found that the $^{214}$Po
half-life averaged at 480 days is equal to $163.45\pm0.04$~$\mu$s.
\vskip 10mm

\section{\label{sec:intro}Introduction}

At the last time in works intended to search for limits of the
realization of the decay constant conservation law, a level of
sensitivity not less than $2\times10^{-4}$  was reached for several
radioactive isotopes. In the work \cite{a1} the authors showed an
amplitude of a possible annual variation of  the $^{198}$Au
half-life ($T_{1/2}=2.69445$ days), that was measured with the
relative uncertainty of  $\pm7\times10^{-5}$ does not exceeds
$\pm2\times10^{-4}$ of the central value. Variations with periods
from several hours up to one year were excluded at the level of
$9.6\times10^{-5}$ (95\% C.L.) during the measurements of the
$^{137}$Cs half-life ($T_{1/2}=10942$ days) in the Ref. \cite{a2}.
The annual variation was excluded at the level of
$8.5\times10^{-5}$ (95\% C.L.). Variations of an activity with
periods of 3-150 days were excluded at the level of
$2.6\times10^{-5}$ (99.7\% C.L.) during the measurement of the
$^{40}$K activity in the Ref. \cite{a3}. It was shown that an
amplitude of the annual variation does not exceeds of
$6.1\times10^{-5}$ (95\% C.L.). Variations of an activity with
periods less then one year were excluded at the level of
$4\times10^{-5}$ during the measurement of the $^{232}$Th activity
in the Ref. \cite{a3} too.

A count rate of the detector recording the source radiation was a
subject of investigations in the all mentioned works. A high
sensitivity of the measurements was reached by using of a
relatively high count rate ($\sim10^3$ s$^{-1}$), of a control and
a stabilization of conditions of the measurements and by an use of
additional  arrangements  for a shield of the set-ups against of
outer background.

%%%%%%%%%%%%%%%  1

The reached limitations are multiply exceed amplitude values of
$\sim1.5\times10^{-3}$ of the $^{32}$Si and $^{226}$Ra count rates
of annual variations discussed in the work \cite{a4}. The authors
examined possibilities of an appearance of such variations as a
result of seasonal variations of the detector characteristics or
of  the one of an annual modulation of the isotope decay rates
themselves under the action of an unknown factor depending of the
Earth-Sun distance. It is obviously that any conclusions about a
possible new physical effect could be made only after  complete
exclusions of variations caused by the influence of the known
terrestrial geophysical, climatic and meteorological factors on
source-detector couple count rates.

Unfortunately, not all such factors could be detected and be taken
into account during the measurement and  data processing. For
example, an annual variation with the amplitude of
$(4.5\pm0.8)\times 10^{-5}$ was found as a result of a processing
of collected during 500 days  data  sample of Earth's surface
measurement with $^{40}$K source in the Ref. \cite{a3}. It was
found that this variation corresponded completely to the known
annual variation of the cosmic rays intensity and could be
explained by a cosmic rays background event contribution to the
total detector count rate. A variation with the $\sim300$ days
period and of $4\times 10^{-5}$ amplitude was found in the data
collected during 480 days in the underground measurement with the
$^{232}$Th source.  It was found that this variation correlated
with a variation of  a daily averaged dead time per event and
could be explained by a modulation of the RC circuit providing the
shaping time of the amplifier.

%%%%%%%%%%%%%%%%%% 2

The weak point of the experiments intended to monitor a stability
of a controlled radiation count rate is their high sensitivity to
the similar variations of  measurement conditions. It seems that
this shortcoming becomes unimportant in the case of the decay
constant determination based on an direct registration of a
nuclear life time between its birth and  its decay.  The same
method was realized by us Ref. \cite{a5} for the $^{214}$Po which
decays with 164.3~$\mu$s half-life \cite{a6} by emitting the 7.687
MeV $\alpha$-particle. This isotope appears mainly in the exited
state ($\sim87$\%) in the $^{214}$Bi $\beta $-decay. Half-lives of
the exited levels does not exceed 0.2 ps \cite{a7} and they
discharge instantly relative to the scale of  the $^{214}$Po
half-life. Energies of the most intensive $\gamma $-lines are
equal to 609.3 keV (46.1\% per decay), 1120 keV (15.0\%) and 1765
keV (15.9\%). So, the  $\beta $-particle and $\gamma $-quantum
emitted at the moment of a birth of $^{214}$Po nuclear form
start-signal and the $\alpha $-particle emitted at the decay
moment forms stop-signal. Measurements of ``start-stop'' time
intervals allow one to construct decay curve at an observation
time and to determine the half-life time from it's shape. The
$^{226}$Ra source ($T_{1/2}=1600$ years) was used as a generator
of $^{214}$Bi nuclei which arise in the decay sequence of the
mother isotope.

The direct measurement of a nuclear life time allows one moreover
to study the radioactive decay law itself. The theoretical models
discussed in Ref. \cite{a8, a9} predict that the decay curves
could deviate from the exponential law in the short- and very
long-time regions of the time scale. The theoretically predicted
\cite{a10, a11, a12} so called quantum Zeno effect consisting in a
slowing down of the decay rate in a case of constant observations
at the decaying object presents a special interest. Experimentally
Zeno effect was found in repeatedly measured two-level system
undergoing Rabi transitions \cite{a13}, but not observed in
spontaneous decays.

At the first stage of our measurements, a  limitation to the
possible annual variation amplitude was set at the level of
$3.3\times10^{-3}$. Factors limiting a sensitivity were revealed
and ways of its optimization were designed.

In the present work the fulfilled improvements of the set-ups, of
measurement methods   and of data processing are described. An
analysis of possible sources of systematic errors is performed and
obtained results are presented.

\section{Method of measurements}

The TAU-1 and TAU-2 set-ups  used in the Ref. \cite{a5} consist of
the two scintillation detectors  D1 and D2 each. The D1 consisted
of two glued discs of a plastic scintillator (PS) with the 18 mm
diameter (\emph{d}) and 0.8 mm thickness (\emph{h}). A thin
transparent radium spot was deposited preliminary in the center of
inner surface of a one disc. The detector D1 records $\beta
$-particles from the $^{214}$Bi decays and $\alpha $-particles
from the $^{214}$Po decays. The massive detector  D2 consisted  of
NaI(Tl) crystals destines   for $\gamma $-quanta detections.

In the TAU-1 set-up it was used a single NaI(Tl) crystal
(\emph{d}=80 mm, \emph{h}=160 mm) as D2. The D1  is placed on the
end of D2. The light collection is fulfilled  from a surface of
the PS disc installed on a Teflon reflector.

In the TAU-2 set-up  it were used  two NaI(Tl) crystals (D2a and
D2b with \emph{d}=150 mm and \emph{h}=150 mm) placed by ends one
to another with the gap of 10 mm. The D1 is placed into a gap
between D2a and D2b. The light collection is fulfilled from a
lateral side of the PS disc installed into deep narrow well with a
reflecting wall.

The measurements were carried out in the underground low
background conditions at the Baksan Neutrino Observatory of the
Institute for Nuclear Researches RAS (BNO INR RAS,  North
Caucasis).  The TAU-1 set-up was located in the underground
laboratory ``KAPRIZ'' at the depth of 1000 meters of water
equivalent in a low background shield made from
Pb(10cm)+Fe(15cm)+W(3cm). The time duration of the measurements
was equal to 1038 days. The value of a half-life time $\tau $
($\tau \equiv T_{1/2}$) deduced  from the analysis of the
integrated decay curve was found to be equal to
($162.73\pm0.10$)~$\mu$s.

The TAU-2 set-up was located in the low background room in the
underground laboratory DULB-4900 \cite{a14} at the depth of 4900
meters of water equivalent within the additional shield made from
Pb(15 cm). The time duration of the measurements was equal to 562
days. The $\tau $-value was found to be ($164.25\pm0.12$)~$\mu$s.

As  one can see  from shown  results,  both  $\tau $-values agree
with the table ones  within 1~$\sigma $ table limit. However, they
differ one from the other at the level of 13~$\sigma $ due to our
statistics. Such large difference indicates a presence of a
systematic error in the experimental results.

The $\tau $-value equal to
[$163.58\pm0.29$~(stat.)$\pm0.10$~(syst.)]~$\mu$s, that is lying
between our values, was measured at the Gran Sasso in a recent
work \cite{a15}.

A possible source of  the supposed error could be a small
difference of sampling frequencies of the two digital oscilloscope
(DO) used for the digital record of pulses. A time duration of
reference square-wave pulses from a high stability pulse generator
was measured by both DO in order to determine a such possible
difference. It was found that difference of measured time
durations is not exceeds $\pm1$ time-channel for the 3400
time-channels pulse or the uncertainty is not more than
$3\times10^{-4}$.

Later  both set-ups were considerably modernized to improve the
sensitivity.

In the TAU-1 set-up the single-crystal detector D2  was
substituted for the two-crystal detector, that is similar to one
used at the TAU-2. The PS discs in the D1 detector  were replaced
by  two silicon surface-barrier semiconductor detectors (SiD) with
the diameter of a sensitive region of 25~mm. In the preliminary
measurements it was found that the SiD lost forever its working
characteristics at the $\sim0.5$ year if an active spot of the
source was deposited on the window surface in the way described
earlier. The source was made as separate hermetical bag composed
of the two plastic film discs with each thickness of 2.5~$\mu$m
and with the active spot deposited on the inner surface of the one
disc. The source was installed hermetically between the two SiD.

In the TAU-2 set-up just a similar source was installed between
the fresh PS discs in the detector D1 because of a scintillation
characteristic degradation of the PS under the active spot as it
was also found at the first stage of measurements. The subsequent
studies proved that the detection property degradations of the
D1-detectors in the both set-ups were excluded. An activity of
$^{226}$Ra was equal to 50~Bq in the each source.

Registrations of the pulses in both set-ups are carried out by the
two- channel digital oscilloscope LA-n20-12PCI which is inserted
into a personal computer (PC). Pulses are digitized with 6.25 MHz
frequency (160~ns/channel). The DO-pulse registration starts by a
signal from the D2 which detected  $^{214}$Bi decay
$\gamma$-quanta. A D2 signal opens a record of a pulse sequence
within time window of 655.36~$\mu$s   in which  the  first time
interval of  81.92~$\mu$s  represents an event ``prehistory'' and
the last period of 573.44~$\mu $s  represents an event
``history''. Duration of a ``history'' exceeds the three
$^{214}$Po half-lives. The digitized pulses in the TAU-1 recorded
into the PC memory on the whole. Total  count  rate  was  equal to
$\sim 6$~s$^{-1}$. A daily information volume was equal to $\sim
10$ Gb.

In the TAU-2 set-up  an each detected event  is analyzed by the
``on-line'' program. A number of pulses and their time delays are
defined for an each event. ``Wrong'' events are excluded. Only
appearance times and  amplitude values of pulses attributed to the
``right'' events were recorded in the PC memory. A complete
information for the each event allows  to process data in the
different pulse amplitude regions in the sequel. A count rate of
the ``right'' events was equal to $\sim 12$~s$^{-1}$. A rate of
the information accumulation was equal to $\sim 25$
Mb$\times$day$^{-1}$.
%~~~~~~~~~~~~~~~~~~~~~~~~~~~~~~~~~~~~~~~~~~~~  2

\section{Results of measurements at TAU-1.}

\begin{figure}[ht]
\begin{center}
\includegraphics*[width=5.35in,angle=0.]{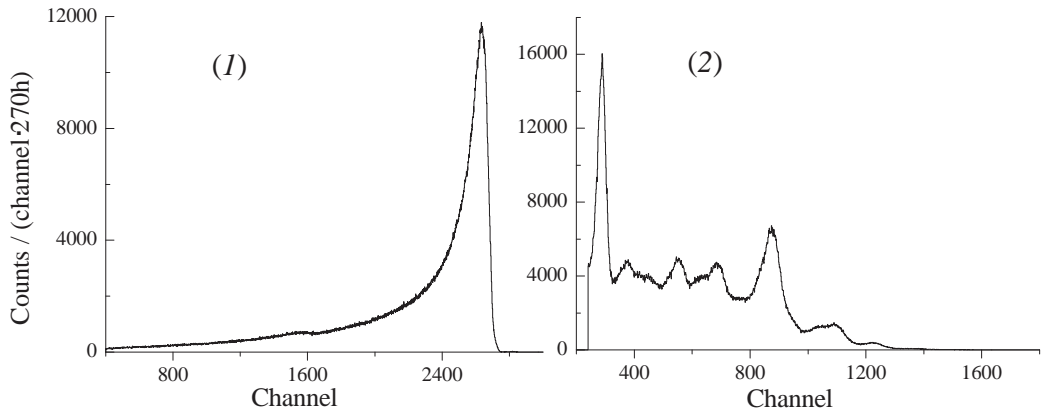}%
\caption{\label{f1} Spectra of the coincided in 573.44 $\mu$s D1
(spectrum \emph{1}) and D2 (spectrum \emph{2}) pulses collected by
the TAU-1 set-up at 270 hours. }
\end{center}
\end{figure}

Spectra of coincident pulses collected by the TAU-1 set-up during
270 hours are shown in Fig.\ref{f1}.
The spectrum \emph{1} represents the data from the detector D1 and
the spectrum \emph{2}  is the data from the detector D2.  An
amplitude threshold value of triggering pulses in the D2 channel
corresponds to  450 keV.  The main peak of the spectrum \emph{1}
conforms to the 7.69 MeV $\alpha$-particle from the $^{214}$Bi
decays. The peaks from decays of the $^{226}$Ra (4.79 MeV) and
$^{222}$Rn (5.49 MeV) are also presented at the spectrum. They
were formed mainly as a result of random coincidences of the D2
background pulses with amplitude values above the threshold one
and of the D1 $\alpha$-pulses with the corresponding energy that
came into the frame of the window duration.

The D2 background is created by a residual radioactivity of the
shield and of the detector materials and  also by  $\gamma$-quanta
from decays of the $^{214}$Pb appearing in the $^{226}$Ra decay
sequence. Random coincidences form a temporal-uniform background
that is situated under the exponential decay curve which is
produced by the true delayed $\gamma$-$\alpha$ coincidences from
decays of the $^{214}$Po nuclei. A small part of random
coincidences is created by the $^{214}$Po $\alpha$-pulses that
have no accompanying  $\gamma$-pulses in those cases when a
$^{214}$Po nucleus  appeared as a result of a $^{214}$Bi
$\beta$-decay on the ground level ($\sim 13$\%),  or in the  cases
when the normal $^{214}$Po decay was delayed for a time more then
the time window duration.

A spectrum of the $^{214}$Bi $\beta$-particle energy-depositions
in the SiD makes up a  small part of the $^{214}$Po
$\alpha$-particle energy-depositions because of an absence of a
pulse amplitude dependence on   particle types. The
$\beta$-particle pulses and $\alpha$-particle pulses are
overlapping at short delay times, and in the result amplitude
values of summarized pulses  will fall into the right slope of the
$\alpha$-peak in the spectrum. Owing to this reason, the decay
curve constructed for the narrow energy region corresponding to
this slope will be enriched in the events with the short delay
times in comparison with the lower energy events. These decay
curves will show different values of the $\tau$. The $\tau$-value
will depend on the low delay times and of cut threshold value of a
decay curve at a process of  its fitting  by an approximation
curve.  It seems that difference of the found $\tau$-values could
be unessential at the appropriate value of the threshold.

\begin{figure}[ht]
\begin{center}
\includegraphics*[width=5.25in,angle=0.]{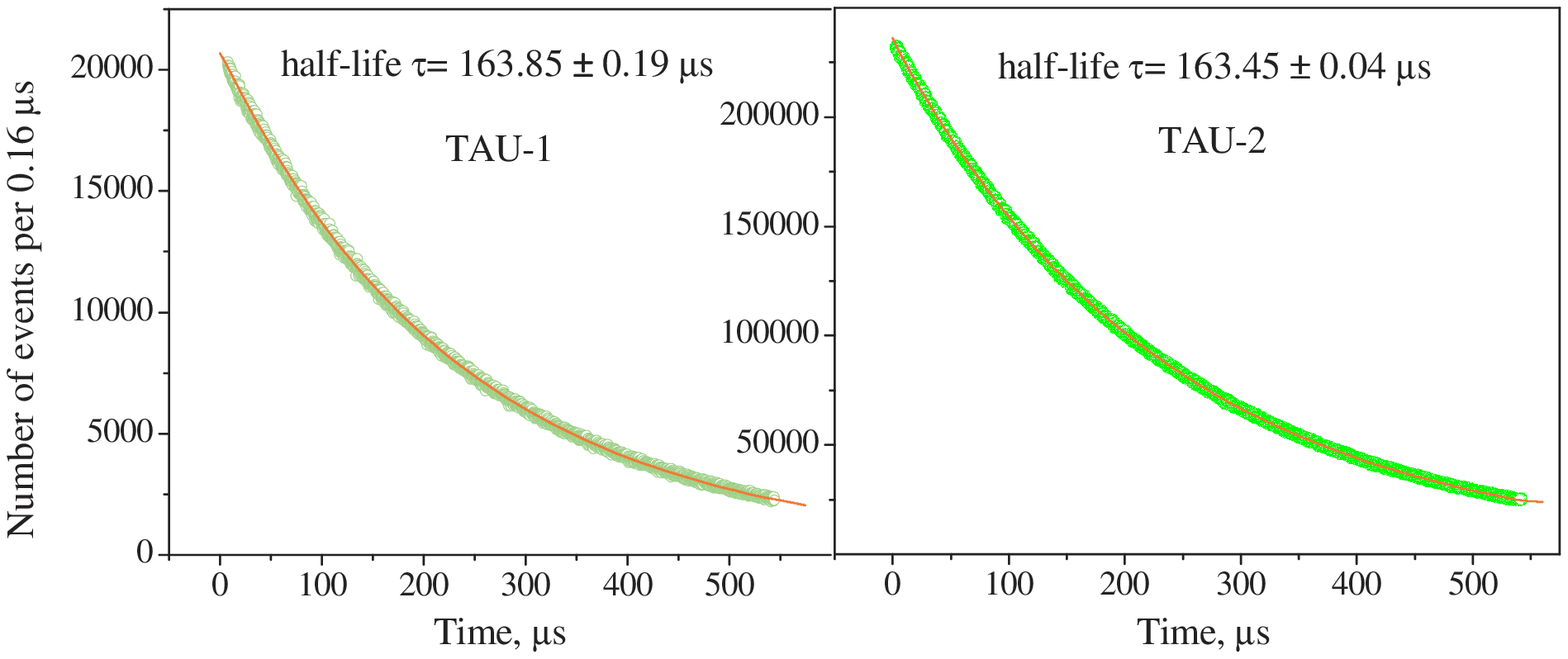}%
\caption{\label{f2} Left panel - a start-stop delays distribution
obtained in 116 days at the TAU-1.
%($\tau=163.85 \pm 0.19$~$\mu$s).
Right panel - a decay-curve of the $^{214}$Po obtained in 480 days
at the TAU-2.
% ($\tau=163.45 \pm 0.04$~$\mu$s).
}
\end{center}
\end{figure}

In Fig.\ref{f2} (left panel) it is shown a start-stop delay
distribution
obtained as a result of a processing of the data collected in 116
days at the TAU-1 set-up. In addition at these measurements the
TAU-1  was shielded with Pb(15~cm)+Cu(8~cm) and was situated near
the TAU-2 set-up in the DULB-4900. An algorithm of the "off-line"
program provided the determinations of the maximum positions
$t_{1m}$ and $t_{2m}$ of pulses from D1 and from D2, used for the
determinations of the delay times. The following step put into
operation a correction on the pulse front durations $t_{1f}$ and
$t_{2f}$ in order to find start points of the pulses.

\begin{figure}[ht]
\includegraphics*[width=5.35in,angle=0.]{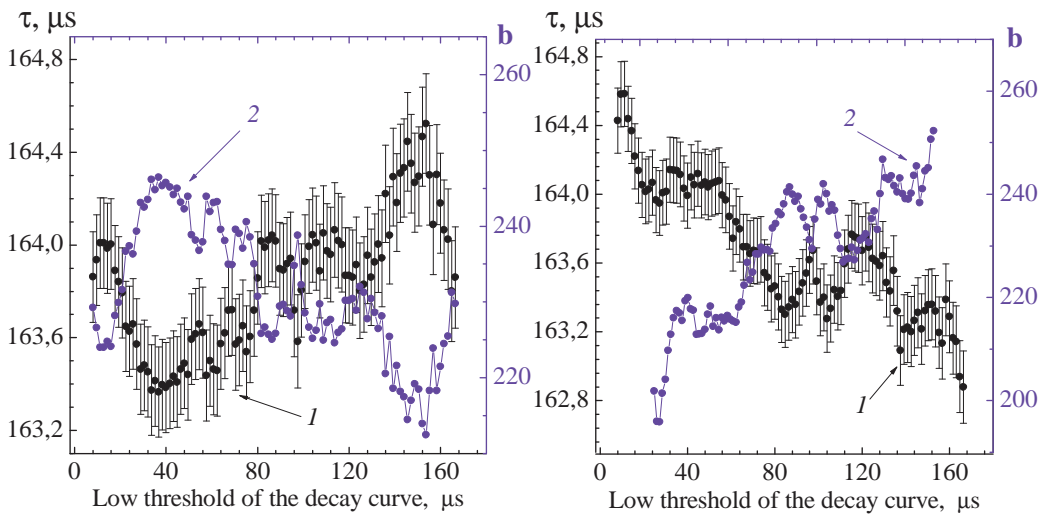}%

\caption{\label{f3} Dependences of a half-life $\tau$-value (curve
\emph{1}, left scale) and a background base amplitude \textbf{b}
(curve \emph{2}, right scale) on  a decay curve cut threshold
obtained with the TAU-1 in the DULB-4900 (left panel) and in the
KAPRIZ (right panel).}

\end{figure}

A value of the delay time $\Delta t$  was calculated as $\Delta
t=(t_{1m}-t_{1f})-(t_{2m}-t_{2f})$. The delay times distribution
shown in Fig.\ref{f2} is approximated by an exponential function
in the form
\begin{eqnarray}\label{eq1}
  y = \mathbf{a} \times \exp [ - \ln (2) \times t/\mathbf{\tau}] +
  \mathbf{b}
\end{eqnarray}
by means of variations of the parameter values \textbf{a},
\textbf{b} and of \textbf{${\mathbf \tau }$} according to the
algorithm of the minimum $\chi$-square method. A value of the
half-life time \textbf{${\mathbf \tau }$ } is found from this
approximation. The approximations were repeated many times using
different threshold cuts to reveal a possible influence of the
$\beta$-particle pulse in the D1 channel on a determination
accuracy of the $\gamma$-$\alpha $ delay time measurements. The
dependences of the obtained $\tau $-values on a cut threshold ones
(curve \emph{1}, left scale) and on the corresponding
\textbf{b}-values of background contributions  (curve \emph{2},
right scale) are shown in Fig.\ref{f3} (left panel).
They looks like strictly anti-phased  at the used right vertical
scales.

In the present experiment before data recording, an event
selection realized strictly under the requirement \textbf{in the
presence of one and only one event within time window of $0\div
560$~$\mu$s of "history"}. In accordance with statistical laws
\cite{a16}, a distribution of random coincident events (under this
selection rule) will be uniform within chosen time interval. This
means that an amplitude value of the background contribution does
not depend on the delays between the D2 and D1 pulses. So, the
observed variations of \textbf{b}-values reflects only ratio
changes of exponent-background for concrete part of decay curve
due to using formal approximation rules.  In the other words, a
shape of the decay curve does not described by a single exponent
within the time region of 10-30 $\mu$s at a level of an accuracy
of $\sim 0.4$\%.

A more precise value of the $\tau$ could be found in a case when
the amplitude b of the background contribution would be known.
Such possibility was found at the present phase of the experiment
during the data analysis.

The method is consists on the measurement of the delays between
the D2 start pulses and the preceded D1 pulses within settled time
interval, that is from the ``prehistory''. A distribution of the
delay times for such random events will be the same as for the
random event in the ``history''. Unfortunately, in the before
collected data a determinations the preceded delays were
impossible  because of  the ``on-line'' PC treatment program of a
preliminary event selections excluded such events before the
recording. At present time this forbidden is eliminated  and all
TAU-1 events are collected.

In principle, a difference of  the $\tau$-value measurements
obtained at the TAU-1 and TAU-2 set-ups could be explained by the
other way. The different gravitation potentials at the KAPRIZ and
DULB-4900 laboratories could be a reason of such a difference, if
there is any dependence of the $\tau$-value on values of the
gravitation potentials at  places on the Earth where the set-ups
are situated. A gravity force in the DULB-4900 is less at
$1\times10^{-4}$ than the one in the KAPRIZ laboratory due to the
gravitation of the rock mass above the deep laboratory as it was
measured in the Ref. \cite{a17}. This difference is much more than
the periodic variations of the Earth's surface gravitation
potential caused by the Sun or  by the Moon orbital moving.

In order to check this assumption, the TAU-1 set-up was replaced
from the DULB-4900 laboratory into the KAPRIZ one.  The conditions
of the measurements and data processing were kept in the same way
by chance. Time of the data accumulation was equal to 88 days. The
results, that are similar to data shown in  Fig.\ref{f3} (left
panel), are presented in Fig.\ref{f3} (right panel). One can find
from a comparison of the data in regions of minimums of the
$\tau$-values at 24~$\mu$s that the $\tau$-value is equal to
$163.5\pm0.2$~$\mu$s in the DULB-4900 place and is equal to
$164.0\pm0.2$~$\mu$s in the KAPRIZ place. The results are in
agreement  within  the 1.5~$\sigma$ interval.

\section{Search for long duration variations of the TAU-2 data}

\begin{figure}[ht]
\begin{center}
\includegraphics*[width=3.5in,angle=0.]{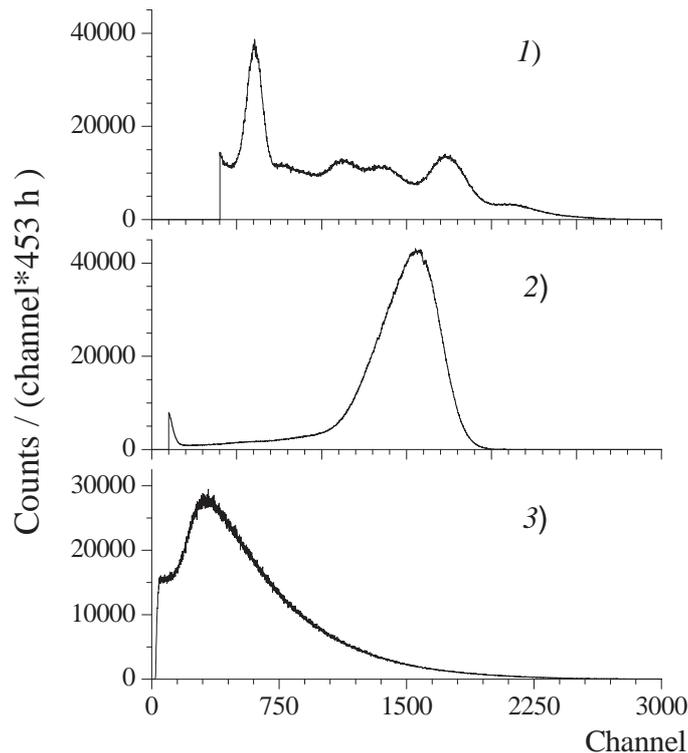}%
\caption{\label{f4} Spectra of the delay coincident  D1 (spectrum
\emph{2}) and D2 (spectrum \emph{1}) pulses and the spectrum
\emph{3} of the first D1 pulses collected by the TAU-2 set-up at
435 hours.}
\end{center}
\end{figure}

A scintillation detector D1 in the TAU-2 has a relative $\alpha
/\beta $ light output equal to $\sim0.1$ \cite{a18}. Because of
it, the pulses from $\alpha$-particles and $\beta$-ones have the
comparable amplitudes. This circumstance was used to a preliminary
selection of the ``useful'' events by the ``on-line'' program
prepared the data for a PC recording. The record program selects
only the events with the two pulses  in the D1 channel. The first
of them ($\beta $) is in a prompt coincidence and the second one
($\alpha $) is in a delayed coincidence  with the start pulse
($\gamma $) in the D2 channel. The corresponding spectra of the
$\beta$-pulses (spectrum \emph{3}) and $\alpha$-pulses (spectrum
\emph{2}) from the D1 detector and of the $\gamma$-pulses
(spectrum \emph{1}) of the D2 detector collected at 435 hours are
shown in Fig.\ref{f4}.
The peak at the channel $\sim1550$ on the spectrum \emph{2} is
formed by the 7.69 MeV $\alpha$-particles. The total time of the
data collection is equal to 480 days in the period of October 2012
-- January 2014.

\begin{figure}[ht]
\begin{center}
\includegraphics*[width=1.85in,angle=270.]{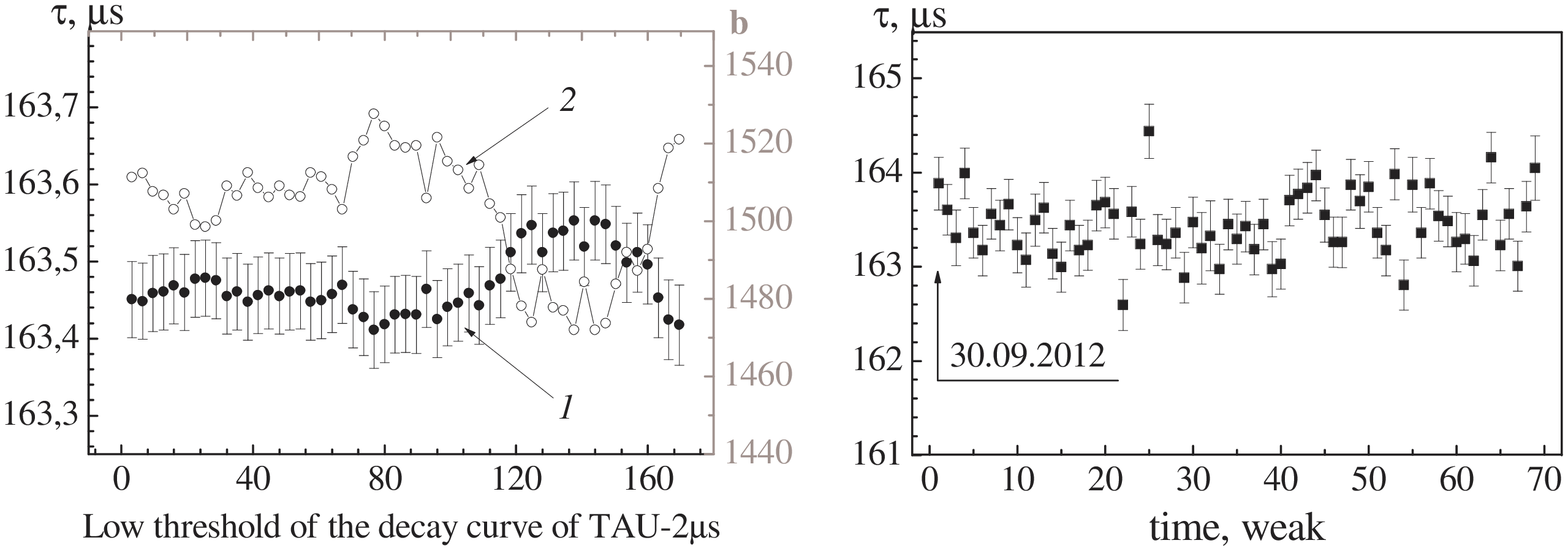}%
\caption{\label{f7} Left panel - dependences of a half-life
$\tau$-value (curve \emph{1}, left scale) and a background base
amplitude \textbf{b} (curve \emph{2}, right scale) on a decay
curve cut threshold obtained with the TAU-2. Right panel -
dependence in time of the $\tau$-value obtained at the TAU-2 with
the week step.}
\end{center}
\end{figure}

A decay curve constructed for the total data set is shown in
Fig.\ref{f2} (right panel). The dependences of the $\tau$-value
measurements (curve \emph{1}, left scale) and of the defined
\textbf{b}-parameter values (curve \emph{2}, right scale) on the
cut threshold values are presented in Fig.\ref{f7} (left panel).
A value of the $\tau$ is equal to $163.45\pm0.04$ at the threshold
of  24~$\mu$s. It is compatible with the $\tau$-value within
1~$\sigma$ interval, that was found in the DULB-4900 experiment
with the TAU-l.

The constant linear contributions of 500, 1000 and 1400 were
subtracted from the decay curve data  to test a dependence of the
procedure for the \textbf{a}-, \textbf{b}- and $\mathbf{\tau}$-values
definitions on the background contributions. The \textbf{a}-,\textbf{b}-
and $\mathbf{\tau}$-values were determined with the standard
procedure for each new decay curve. The  \textbf{a}- and
$\mathbf{\tau}$-values were found to be the same for all three background
contributions, and the b-values was reduced exactly by the
subtracted constants.

This means that: 1) a shape of a background contribution is really
flat  since  in the opposite case the parameters of the exponent
should be changed to compensate an increased contribution of  the
background nonlinear part; 2) an accuracy of the separation of the
experimental decay curve form by the exponential part and by the
flat one does not depend on the background values in the treated
limits.

The time-continuous data set was divided to the equal duration
time intervals to search for possible time variations. The decay
curves have been constructed for each data portions and the
corresponding $\tau$-values have been defined. So,  the continuous
time-interval sequence of the $\tau$-value measurements within the
specified time step has been found.  The dependence of the found
$\tau$-values  on time with a week time step of the distribution
is shown in Fig.\ref{f7} (rigth panel). The $\tau$-values were
defined by means of a $\chi^2$-approximation of the decay- curves,
each collected during 7 days, by an exponential function
(\ref{eq1}). The used time window of delays was $3.2 \div
560$~$\mu$s.

There is the statistically more powerful maximum likelihood method
for estimation of exponent parameters in experimental data
treatments, but in order to use it a value of the background
contribution b should be  determined by means of an independent
direct measurements or of any additional data analysis.

\begin{figure}[ht]
\begin{center}
\includegraphics*[width=5.25in,angle=0.]{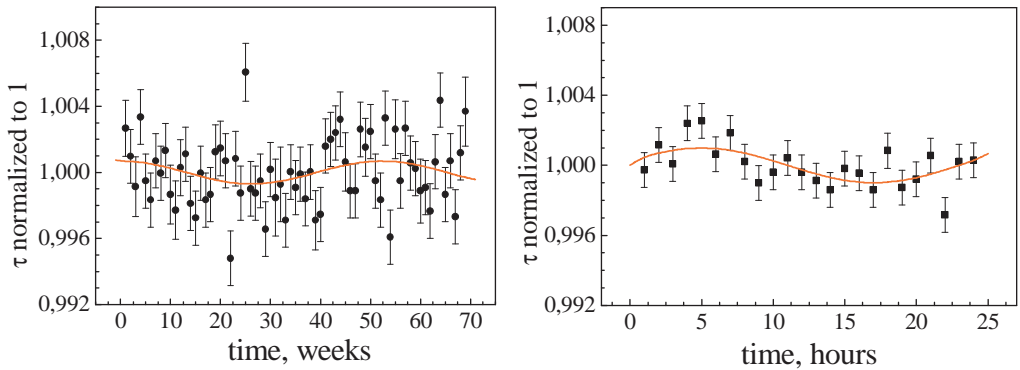}%
\caption{\label{f10} Left panel - dependence in time of the
normalized on average $\tau$-value obtained at the TAU-2 with the
week step (black points) and an approximation function (\ref{eq2})
$f(t)=1+6.9\times10^{-4}\times sin\{2\pi/365\times (t+93)\}$ (colour
curve). Right panel - dependence with 1 hour step of the
$\tau$-value on solar day time, colour curve - $f(t)=1+1\times
sin\{2\pi/24\times (t+1)\}$.}
\end{center}
\end{figure}

To search for a possible annual variation of the found
$\tau$-data, they were normalized to the averaged  values and were
compared with a periodical function
\begin{eqnarray}\label{eq2}
  f(t,\varphi)=1+A \times sin\{\omega (t+\varphi )\},
\end{eqnarray}
where $\omega=2\pi /365$ days$^{-1}$, \emph{t} -- day of year,
\emph{A} -- an amplitude of the variation, $\varphi $ -- a phase
shift relative to the 1 January. Here \emph{A} and $\varphi$ are
used as trial parameters for to find best fit. The
$\varphi$-parameter was varied from 1 to 365 with the step of
1~day. A correlation coefficient $k(\varphi)$ between $\tau$-value
sequence and $f(t,\varphi)$ was calculated for the each
$\varphi$-value. The maximum value $k=0.23$ has been reached  at
$\varphi=90\pm10$ days. So, the $\varphi$-phase value of the
periodical function was found. Then a choice of A-value
corresponding to the $\chi^2$ minimum was done and was  found to
be $A=(6.9\pm3.0)\times10^{-4}$. A maximum of the $f(t,\varphi)$
has achieved on the 22 September. The corresponding dependences
are shown in Fig.\ref{f10} (left panel).

The other natural periodic variations exist which are connected
with the rotation of the Earth around its axis. In particular,
oscillations with periods of 24 hours  in the Sun's time. Siderial
time are related to such phenomena. A $\tau$-values sequence
obtained for the one hour step decay curves putting in the 24
hours Sun's day averaged from the TAU-2 data collected during 16
month is shown in Fig.\ref{f10}.
The normalized sequence was approximated by the expression
(\ref{eq2}), where $\omega = 2\pi/24$~h$^{-1}$. A maximum
correlation value of $k=0.6$ was found for the $\varphi=1.0 \pm
0.5$ hours, $A=(10.0\pm2.6)\times10^{-4}$ at the $\chi^2=1.49$ for
$N = 23$. The corresponding dependences are shown in Fig.\ref{f10}
(right panel). There are no variations in the similar data
sequences treated in the sidereal time which exceed statistical
dispersions with the values higher than $A\leq 3\times 10^{-4}$
(90\% C.L.).

\section{ Results and discussion}

The fulfilled modernization of the TAU-1 and TAU-2 set-ups allowed
us to improve considerably a stability of the results and a
sensitivity of long duration measurements of a half-life value of
the $^{214}$Po decay  as it follows from the description given
above. An assumption that obtained in Ref. \cite{a5} $\sim 0.9$\%
difference of the $\tau$-values measured at the old-version of
TAU-1 and of TAU-2 set-up versions could be connected with the
difference of the calibrations of the used digitizers was examined
by means of measurements of the stable rectangle pulse durations.
It was found that an accuracy of the DO calibrations was not worse
than $3\times10^{-4}$. A hypothesis about possible correlation of
$\tau$-values measured by the old TAU-1 and TAU-2 and  of values
of gravity in the corresponding underground laboratories was
tested. The values of a gravitational acceleration in  used two
laboratories are differ at $1 \times 10^{-4}$. Firstly, a
measurement was done with the TAU-1 set-up in one laboratory and
than it was repeated in the other one. A weak dependence of  the
measured $\tau$-values on  decay curve cut thresholds in the low
delays region was obtained in the both series. The changes of the
$\tau$-values do not exceeded 3~$\sigma$  or  $3.7\times10^{-3}$
for the $3.2 \div 30$~$\mu$s  threshold changes at the achieved
statistical level. A difference of the $\tau$-values at the
threshold of 30~$\mu$s does not exceeds a value of
$3.7\times10^{-3}$ and this difference is much lower than the one
measured in Ref. \cite{a5}. A statistic of measurements in the
each laboratory should be increased to a further improvement of
the estimation accuracy.

A dependence of a $\tau$-value on a decay curve cut threshold
could be explained by small distortions of the exponent form at
low delay times due to mistakes of the ``off-line'' program in
processes of determinations of  real delays in cases of front
overlapping of $\beta$-pulses and of $\alpha$-pulses. The
$\tau$-value variations at the time delay above $\sim 60$~$\mu$s
could be connected with statistical deviations of the exponent
form since the statistical weights of the such deviations changed
at the threshold growth. The approximation program redistributes
the exponent part and background line contributions in accordance
with these weights. Variations of the $\tau$-value and the
background contribution \textbf{b} are in anti-phase in accordance
with expression $\Delta \tau / \tau \approx  -0.037 \times \Delta
\mathbf{b}/\mathbf{b}$  as it seen in Figs. \ref{f3} and \ref{f7}
(left panel).

It is evidently  that contribution \textbf{b}  is really constant
under any part of the exponent. A steadiness on the
\textbf{b}-values of the $\chi^2$-algorithm of ORIGIN used for a
decay-curve division between  the  the exponent part and
background line part was tested by comparing of the results
extracted by the program in the processing of the decay curves
obtained as results of constant line subtractions from the primary
decay curves. The identical exponent parameters were found in all
tested cases. The background values \textbf{b}  were found to be
equal to the difference of the primary \textbf{b}-value and
subtracted constant values. This observation gives a certainty in
the interpretation of the data time sequences with different time
steps.

The uncertainties of interpretations could arise on account of
possible independent changes of the exponent and of the background
parameters  during a chosen time step under influences of the
external reasons. The values of the parameters could be found by
different ways, in principle. First, the $\tau$-value for a long
time interval could be calculated as  average of the $\tau$-values
for  short component time intervals. Second, the parameters could
be obtained from the decay curve collected during whole analyzed
time-interval by using of a standard approximation procedure.
Third, the background contribution b could be excluded from a
number of variable parameters before the approximation procedure
by using of a constant value obtained by means of a normalization
of the total data set \textbf{b}-value to the number of events  in
the analyzed time interval. A value of a standard deviation will
decreased in accordance with the ratio of the time step duration
to the total measurement time duration. An adequacy of these
approaches was examined for the week step data decomposition. It
was found that all methods gave similar results but different in
details.

An independent measurement of the background contribution value
should be done to obtain the result unambiguity. It was aware of
possibility of realization of such measurements by using of delays
between random coincide  D1-pulses from ``prehistory'' and
starting D2-pulses. Another possibility could be realized by
admixing of seldom specially prepared stable generator pulses to
the flow of the real pulses. An admixture could be done by direct
input of the pulses into the electronic chain or by the pulse
lighting of the PMTs.

The amplitude and phase in the annual variation of the
$\tau$-value sequence with the week  data step was found to be
$A=(6.9 \pm 3.0) \times 10^{-4}$ and $\varphi = 93 \pm 10$~days.

In order to check a possible occurrence of cyclical variations
with twenty-four hour period,  the  total data set collected at
the 16 months was  transformed to the 24 hour data set by summing
of the information within the same numbered hour in the repeated
24 hours fragmentation. The transformation was repeated for solar,
lunar and stellar times. The twenty-four hour variation with the
amplitude $A=(10.0 \pm 2.6) \times 10^{-4}$ and phase
$\varphi=1\pm0.5$ hour was found in the studied sequences averaged
in the solar time.

There were no any day variations with amplitudes exceeding a
statistical data straggling ($A \leq 3 \times 10^{-4}$ at 90\%
C.L.) in the sidereal time.

It is possible to suppose that the annual variation and the
twenty-four hour one can  have a common origin. In principle, the
found effect of the decay constant variation could be created by
variations of the DO sampling frequency; by variations of the
delay times of the D1 and D2 pulses due to possible variations of
the PMT's time characteristics under the action of the Earth
magnetic field variations; by an unknown physical effect
synchronized with the day Earth circulation and with the annual
one.

An instability of the DO characteristics could be created by a
noticeable changing of an environmental temperature.  However, due
to the continuous monitoring it was shown that a temperature in
the TAU-2 compartment is constant within the limits of
$26.5\pm0.2$~C$^\circ$. This means that the temperature variations
should be excluded from a list of possible reasons of the
$\tau$-variations.

Variations of the supply powers were not considered as an
instability source because of all electronic systems feed by
stabilized voltages. So, by the discussed reasons it seems not
likely  that DO characteristics instability could be a reason of
the  observed  $\tau$-variations.

Possible influences of the Earth's magnetic field variations to
the PMT's characteristics is supposed  to be investigated in the
nearest future.

In the result of the treatment of the whole data sample recorded
by the TAU-2 set-up during 16 months,  the new  value of
$^{214}$Po half-life time averaged over the total observational
period was found to be $\tau = 163.45\pm0.04$~$\mu$s. The value is
compatible with the values from other measurements. Using this new
value of decay constant $\tau$, it is necessary to take into
account the mechanisms shown above.

\section{Conclusions}

The results of analysis of the data obtained with the upgraded
TAU-1 and  TAU-2 set-ups at the new step of measurements are shown
in the presented work. The set-ups are intended to carry out a
long-term control of  the $^{214}$Po half-life constant value. It
is shown that the constant feels the twenty-four hour variation
and the annual one of an unknown nature. The measurements are in
progress.


\begin{thebibliography}{99}

\bibitem{a1}  \textit{Hardy J.C., Goodwin J.R. and Iacob V.E.} //
              Appl.Radiat.Isot. 2012. V.70. P.1931; arXiv:1108.5326 [nucl-ex].
% Hardy, J.C. et al. Appl.Radiat.Isot. 70 (2012) 1931-1933 arXiv:1108.5326 [nucl-ex].

\bibitem{a2} \textit{Bellotti E. et al.} //
             % E. Bellottia, C. Brogginib, G. Di Carloc, M. Laubensteinc, R. Menegazzob
            Phys. Lett. B. 2012. V.710. P.114; arXiv:1202.3662 [nucl-ex].
% Bellotti, E. et al. Phys.Lett. B710 (2012) 114-117 arXiv:1202.3662 [nucl-ex].

\bibitem{a3} \textit{Bellotti E. et al.} //
            % E. Bellotti, C. Broggini, G. Di Carlo, M. Laubenstein, R. Menegazzo, M. Pietroni
            "Search for time modulations in the decay rate of $^{40}$K and $^{232}$Th and influence of a scalar field from the
            Sun." arXiv:1311.7043 [astro-ph.SR]

\bibitem{a4} \textit{Jenkins J.H. et al.} //
             Astropart. Phys. 2009. V32. P.42; arXiv:0808.3283 [astro-ph].

\bibitem{a5} \textit{Alexeyev E.N. et al.} //
             Astropart. Phys. 2013. V46. P.23; arXiv:1112.4362 [nucl-ex].

\bibitem{a6} \textit{S.-C.~Wu} //
             Nucl. Data Sheets 2009. V110. P.681.


\bibitem{a7} {Table of Isotopes}, Seventh Edition, Edited by Firestone R.B. et al., 8th ed. {Willey, New York 1996}.

\bibitem{a8} \textit{Gavriljuk Ju.M. et al.} //
            Nucl. Instr. Meth. A 2013. V729. P.576; arXiv:1204.6424 [physics.ins-det].


\bibitem{a9} \textit{Gopych P.M. and Zaljubovsky L.I. } //
             Fiz. Elem. Chast. Atom. Yadra. 1988. V.19. P.785.


\bibitem{a10} \textit{Fonda L. et al.} //
              Rep. Prog. Phys. 1978. V41. P.587.

\bibitem{a11} \textit{Khalfin L.A.} //
              Physics-Uspekhi 1990. V160(10). P.185.

\bibitem{a12} \textit{Misra B. and Sudarshan E.C.G.} //
              J. Math. Phys. 1977. V18. P.756.

\bibitem{a13} \textit{Facchi P. and Pascasio S.} //
              J. Phys. A: Math. Theor. 2008. V41. P.493001.

\bibitem{a14} \textit{Itano W. et al.} //
              Phys. Rev. A. 1990. V.41. P.2295.

\bibitem{a15} \textit{Bellini G. et al. (BOREXINO Collaboration)} //
              Eur. Phys. J. A  2013. V.491. P.92;
              arXiv:1212.1332v1 [nucl-ex].
\bibitem{a16} Stochastic processes. Doob J. L. Wiley, New York 1953.

\bibitem{a17} \textit{Medvedev M.N.} //
              "\emph{Report from gravity works in the Baksan Valley in 2013}". Report SAI, Moscow, November
               2013.

\bibitem{a18} \textit{Yushkin V.D. and Dzeranov B.D.}
              Scintillation Detectors. M. Atomizdat, 1977.

\end{thebibliography}
\end{document}